# Quantitative agreement of Dzyaloshinskii-Moriya interactions for domain-wall motion and spin-wave propagation

Dae-Yun Kim[1,☆], Nam-Hui Kim[2,☆], Yong-Keun Park[1,3], Min-Ho Park[1], Joo-Sung Kim[1], Yune-Seok Nam[1], Jinyong Jung[2], Jaehun Cho[4], Duck-Ho Kim[1,5], June-Seo Kim[6], Byoung-Chul Min[3], Sug-Bong Choe[1,★], and Chun-Yeol You[2,★]

[1]Department of Physics and Astronomy, Seoul National University, Seoul, 08826, Republic of Korea.

[2]Department of Emerging Materials Science, Daegu Gyeongbuk Institute of Science and Technology (DGIST), Daegu, 42988, Republic of Korea.

[3]Center for Spintronics, Korea Institute of Science and Technology (KIST), Seoul, 02792, Republic of Korea.

[4]Department of Materials Engineering Science, Osaka University, Osaka, 560-8531, Japan.

[5]Present address: Institute for Chemical Research, Kyoto University, Kyoto, 611-0011, Japan.

[6]Intelligent Devices & Systems Research Group, Institute of Convergence, DGIST, Daegu, 42988, Korea.

☆Equally contributed

★Corresponding author: sugbong@snu.ac.kr and cyyou@dgist.ac.kr

**KEYWORDS**

Dzyaloshinskii-Moriya interaction, domain-wall dynamics, spin-wave dynamics

**ABSTRACT**

The magnetic exchange interaction is the one of the key factors governing the basic characteristics of magnetic systems. Unlike the symmetric nature of the Heisenberg exchange interaction, the interfacial Dzyaloshinskii-Moriya interaction (DMI) generates an antisymmetric exchange interaction which offers challenging opportunities in spintronics with intriguing antisymmetric phenomena. The role of the DMI, however, is still being debated, largely because distinct strengths of DMI have been measured for different magnetic objects, particularly chiral magnetic domain walls (DWs) and non-reciprocal spin waves (SWs). In this paper, we show that, after careful data analysis, both the DWs and SWs experience the same strength of DMI. This was confirmed by spin-torque efficiency measurement for the DWs, and Brillouin light scattering measurement for the SWs. This observation, therefore, indicates the unique role of the DMI on the magnetic DW and SW dynamics and also guarantees the compatibility of several DMI-measurement schemes recently proposed.

The interfacial Dzyaloshinskii-Moriya interaction (DMI) is an antisymmetric exchange interaction between spins mediated by heavy metal atoms [1-3]. It is known that the structural inversion asymmetry in the magnetic system generates a sizeable DMI with an energy $E_{\mathrm{DMI}}$ as given by

$$E_{\mathrm{DMI}} = -\vec{D} \cdot (\vec{M}_i \times \vec{M}_j), \tag{1}$$

where $\vec{D}$ is the DMI vector and $\vec{M}_i$ and $\vec{M}_j$ are the neighboring local magnetization. The DMI-induced antisymmetric exchange interaction has recently received great attention because of its crucial role in spintronic materials, such as the stabilization of chiral magnetic domain walls (DWs) or the formation of the magnetic skyrmion [4-7]. Numerous efforts have been devoted to investigating the role of the DMI on the magnetization process [6-8] and also, various experimental schemes to quantify the strengths of the DMI have been proposed [9-17]. These experimental schemes are mainly based on either the DMI-induced chirality of the DWs [9-14] or the DMI-induced non-reciprocity of the spin waves (SWs) [15-18]. Up to now, however, the measurement results among these experimental schemes have been in conflict [19]. It is not clear yet whether this discordance can be attributed to experimental artifacts or the intrinsic nature of the DMI. For example, for the former case, Kim *et al*. [20] recently demonstrated that the presence of a sizeable additional antisymmetric contribution [21-27] such as chiral damping [22, 23] causes experimental inaccuracy in the DW-speed-asymmetry DMI-measurement scheme [10, 12, 21, 22, 27] and this inaccuracy can be removed by adjusting the measured value by the amount of the DW saturation field [20]. Or, in the latter case, if the DMI has angular dependence on the magnetization, the effective amount of the DMI over the magnetic objects is different for DWs (in which the magnetization rotates 180°) and SWs (which has small angle deviation from uniform magnetization).

To evaluate all these possibilities, here we investigated the compatibility (or incompatibility) of the DMI based on the strength of the DMI, determined using two different magnetic objects, the DWs and SWs. After careful measurement and analysis, and the removal of any possible experimental artifact, we observed that the two magnetic objects exhibit the same strength of DMI, signaling that the angular dependence of the DMI is minimal.

A series of films comprised of magnetic multilayers of Pt/Co/X with different X (= Al, Au, Cu, Pt, Ta, Ti, and W) were prepared with a DC magnetron sputtering system, as shown in Figure 1(a). The films were deposited on Si/SiO$_2$ substrates with a 5.0-nm-thick Ta adhesion layer and a 1.5-nm-thick Pt protection layer [28]. The detailed layer structure was 2.5-nm Pt/0.9-nm Co/2.5-nm X, where the Co layer thickness $t_{Co}$ was chosen to satisfy the optimal experimental conditions. These conditions were: 1) within the clear DW motion phase for the DW-based measurement (as shown in Fig. 1(b); for the other samples, please see Supporting Information I); and 2) as thick as possible for better sensitivity in the Brillouin light scattering (BLS) measurement of the SW dynamics. All the samples exhibited strong perpendicular magnetic anisotropy (PMA).

For the DW-based measurement, we employed the spin-torque efficiency measurement scheme proposed by P. P. J. Haazen *et al*. [9]. We will denote the scheme as the '$\varepsilon_{ST}$-measurement scheme' hereafter and when we compare it with other schemes later. In this scheme, the spin-torque efficiency is measured with respect to an external longitudinal magnetic field and then, the strength of the DMI is estimated from the characteristic variation in the spin-torque efficiency.

For this measurement, continuous film was patterned into a micro-wire structure with electrodes and a writing line using a photo-lithography technique, as shown by Fig. 1(c). The sample was first saturated by applying an external out-of-plane magnetic field larger than the coercive field and then, a DW was created near the DW writing electrode (white vertical line) by injecting a current through it [29]. Then, with the application of current bias with a current density $J$ through the magnetic wire, the depinning field $H_{\text{dep}}$ was measured by sweeping an external out-of-plane magnetic field until the DW moved from the initial position to the probing spot (red circle). By repeating this procedure with changing $J$, the spin-torque efficiency $\varepsilon_{\text{ST}}$ was determined to be $\varepsilon_{\text{ST}} = -\partial H_{\text{dep}}/\partial J$ [9, 29].

Figure 2(a) exemplifies the plot of the measured $\varepsilon_{\text{ST}}$ as a function of the in-plane magnetic field $H_x$ for the X=Ti sample (for the other samples, please see Supporting Information II). As guided by the solid lines, the curve exhibits an inversion anti-symmetry with respect to an axis depicted by the bold vertical line. This anti-symmetric behavior indicates that $\varepsilon_{\text{ST}}$ can be mainly attributed to the spin-orbit torque, rather than the spin-transfer torque [9, 29, 30]. This observation is consistent with Ref. [30], in which a sizeable spin-transfer torque appeared only in the case of a thin ferromagnetic layer (~0.3 nm), while the present samples have a relatively thick ferromagnetic layer (~0.9 nm). Therefore, we used $\varepsilon_{\text{ST}} \approx \varepsilon_{\text{SOT}}$, where the spin-orbit torque efficiency $\varepsilon_{\text{SOT}}$ is given by

$$\varepsilon_{\text{SOT}} = \frac{\hbar \theta_{\text{SH}}}{2 e M_S t_{\text{Co}}} \cos \psi, \tag{2}$$

where $\theta_{\text{SH}}$ is the net spin Hall angle of the system, $M_S$ is the saturation magnetization, and $t_{\text{Co}}$ is the thickness of the ferromagnetic Co layer [9, 31].

According to Ref. [6, 7], the angle $\psi$ of the magnetization (purple arrow) inside the DW is determined by the counterbalance between the DMI-induced effective magnetic field $H_{\text{DMI}}$ (red arrow) and the DW anisotropy field (blue arrow), as depicted in Fig. 2(b). For the case where $H_{\text{DMI}} = 0$, the DW energetically prefers the Bloch-type chirality ($\cos\psi = 0$) due to the DW anisotropy caused by the dipolar interaction, corresponding to $\varepsilon_{\text{ST}} = 0$ from Eq. (2). For the other case, when $H_{\text{DMI}} \neq 0$, the DW deviates from the Bloch-type chirality. For that case, the Bloch-type chirality can be recovered by applying an in-plane longitudinal magnetic field $H_x^*$ to compensate $H_{\text{DMI}}$ i.e. $H_x^* + H_{\text{DMI}} = 0$. Therefore, in Fig. 2(a), the intercept to the $x$ axis (red vertical line) indicates the typical magnitude of $H_x^*$ required for the Bloch-type DW chirality ($\cos\psi = 0$) and thus, one can quantify the strength of $H_{\text{DMI}}$ using the measured value of $H_x^*$ via the relation $H_{\text{DMI}} = -H_x^*$. By repeating this procedure, the strengths of $H_{\text{DMI}}$ were measured for all the samples.

From the measured values of $H_{\text{DMI}}$, one can estimate the DMI strengths $D_{\text{DW}}$ by using the relation

$$D_{\text{DW}} = (\mu_0 \lambda M_S) H_{\text{DMI}}, \qquad (3)$$

with the DW width $\lambda$ ($= \sqrt{A/K_{\text{eff}}}$), where $A$ is the exchange stiffness constant and $K_{\text{eff}}$ is the effective PMA constant. To estimate the values of $K_{\text{eff}}$ ($= \frac{1}{2}\mu_0 H_K M_S$), the $H_K$ values were measured by vibrating a sample magnetometer with a hard-axis configuration. Then, by applying the literature values of $A$ (=2.2×10⁻¹¹ J/m) and $M_S$ (=1.4×10⁶ A/m) for Co [32], the magnitudes of $D_{\text{DW}}$ were finally estimated. The estimated magnitudes of $D_{\text{DW}}$ are listed in Table I together with the measured values of $H_{\text{DMI}}$ and $H_K$ for all the samples with different X.

For a quantitative comparison with the results of the DW-based $\varepsilon_{\text{ST}}$-measurement scheme, the strengths of the DMI were measured again with the SW-based configuration by using the BLS measurement [15-17]. We will denote that scheme as the 'BLS-measurement scheme' hereafter and when we compare it with other schemes.

According to Ref. [15], among the various SW modes the DMI mainly interacts with the surface SW mode—the so-called Damon-Eshbach (DE) mode [33]— depending on the directions of the wave vector and magnetization. In the DE mode, both the wave vector $\vec{k}_{\text{SW}}$ and the magnetization $\vec{M}$ lie in the film plane, and are orthogonal to each other, as depicted in Fig. 3(a).

To produce this situation in the experiment, the magnetization of our PMA samples were turned to the film plane by applying a transverse in-plane magnetic field $H_y$ (=1.3 T) along $-\hat{y}$ direction, sufficiently stronger than the anisotropy field $H_K$. The strengths of $H_K$ ($= 2K_{\text{eff}}/\mu_0 M_S$) are listed in Table I. When $H_y$ is applied along the $-\hat{y}$ direction, the precession of magnetization is in the counter-clockwise direction, as depicted by the curved black arrows. With this precession, two SW modes are allowed, each of which are distinguished by their propagation directions. The cross sectional views of the two modes with opposite wave vectors along the $+\hat{x}$ and $-\hat{x}$ directions are depicted in Figs. 3(b) and (c), respectively.

It is worth noting that, due to the opposite directions of the wave vectors, the two modes have opposite angles between the local magnetization neighboring in space as seen by Figs. 3(b) and (c). Therefore, $\vec{M}_i \times \vec{M}_j$'s are opposite to each other between the two modes, resulting in the opposite signs of $E_{\text{DMI}}$ in Eq. (1) with a fixed $\vec{D}$ in a given film. Therefore, the two modes have DMI-induced energy shifts that are opposite to each other, which results in the non-reciprocal modification of wave vector for a given SW frequency, as shown in Figs. 3(b) and (c).

This non-reciprocity between the SW modes was experimentally verified by Cho *et al.* [15] via the BLS measurement. Fig. 3(d) shows the typical BLS spectrum of a given SW wave vector $k_\text{SW}$ (=0.0167 nm$^{-1}$) for the sample with X = Ti. The figure visualizes the non-reciprocity of the SW modes with opposite shifts $\pm \Delta f_\text{SW}$ in peak frequency for the two modes with the opposite wave vectors $\pm k_\text{SW}$, respectively. According to Ref. [15], $\Delta f_\text{SW}$ is given by

$$\Delta f_\text{SW} = \frac{2\gamma D_\text{SW}}{\pi M_\text{S}} k_\text{SW}, \tag{4}$$

where $\gamma$ is the gyromagnetic ratio and $D_\text{SW}$ is the DMI strength. Fig. 3(e) confirms the linear proportionality between $\Delta f_\text{SW}$ and $k_\text{SW}$ (for the other samples, please see Supporting Information III). The values of $D_\text{SW}$ are then quantified from the linear slope, as listed in Table I for all the samples.

Figure 4 plots the measured values of $D_\text{SW}$ with respect to $D_\text{DW}$ for various X. Great agreement is seen between $D_\text{DW}$ and $D_\text{SW}$. This observation permits us to write: 1) The same analytic form of Eq. (1) is valid for both cases of basically uniform magnetization (corresponding to the SW case) and largely-varying non-uniform magnetization (corresponding to the DW case). Also, 2) the present two measurement schemes give the same results, consistent with each other, which confirms the reliability of the recently-developed diverse measurement schemes and thus, provides a way towards establishing measurement standards.

It is worth comparing the present results with the other widely-used measurement scheme proposed by Je *et al.* [10], which is based on the symmetry of the DW speed with respect to $H_x$. We well denote that scheme as the '$v_\text{DW}$-measurement scheme' hereafter. The $v_\text{DW}$-measurement scheme analyzes the shift in the symmetry axis of the DW speed, which is similar to the $\varepsilon_{ST}$-measurement scheme, which analyzes the shift in the anti-symmetry axis of $\varepsilon_{ST}$ with respect to $H_x$. Both the $v_{DW}$- and $\varepsilon_{ST}$-measurement schemes are applicable <u>*only when the data show symmetry and anti-symmetry with respect to the symmetry and anti-symmetry axes*</u>, respectively. Failure of this criterion results in significant artifacts.

For example, for the DW speed case, recent studies have shown that additional asymmetries exist, such as chiral damping [22, 23] and intrinsic variation in the DW width [24]. These additional asymmetries destroy the symmetric behavior of the DW speed and thus, it is not possible to unambiguously determine the symmetry axis [20], as was confirmed with our present samples (Supporting Information IV).

Kim *et al.* [20] recently demonstrated experimentally how to recover the symmetry axis with the presence of sizeable additional asymmetries. After properly correcting the symmetry axis, both the $v_{DW}$- and $\varepsilon_{ST}$-measurement schemes provided the same results. For the BLS-measurement scheme, due to the possible offset in $\Delta f_{SW}$, it is essential to confirm the linear proportionality between $\Delta f_{SW}$ and $k_{SW}$ for better accuracy of the results.

Finally, we would like to discuss the applicable conditions which are best for each measurement scheme. For intuitive understanding, please refer to Figure 5, which summarizes the following discussions. The $v_{DW}$- and $\varepsilon_{ST}$-measurement schemes maintain their sensitivity to the thin ferromagnetic layers down to a few angstroms [10, 30]. However, since both schemes are based on the DW motion, these schemes cannot be applied to thick ferromagnetic layers that usually show a striped or dendritic phase. This is because in those phases, the minimum DW roughness needed to guarantee uniform DW chirality along the DW is not achieved, and thus a sizeable inaccuracy in DMI strength may occur with the present measurement schemes.

The blue perpendicular dashed line in Fig. 5 indicates the phase boundary between the striped and dendritic phases. In contrast to the case above, the BLS-measurement scheme is more applicable to thick ferromagnetic layers, even to thick films with in-plane magnetic anisotropy. However, the BLS-measurement scheme cannot be applied to too strong $H_\text{K}$ samples, because it requires an applied in-plane field $H_y$ stronger than $H_\text{K}$. This makes it hard to apply the BLS-measurement scheme to a very thin ferromagnetic layer with large PMA samples. The red perpendicular dashed line in Fig. 5 shows the lower limit of the BLS-measurement scheme.

In addition, since the $v_\text{DW}$- and $\varepsilon_\text{ST}$-measurement schemes quantify the $H_\text{DMI}$ directly from the value of the control parameter $H_x$, there is essentially no lower boundary for the measurable $H_\text{DMI}$ values, whereas the upper boundary is limited by the strength of the applicable $H_x$. The blue horizontal dashed line guides the upper limit of the practically available $H_x$. On the other hand, since the BLS-measurement scheme quantifies the DMI by detecting the frequency shift, the lower boundary of the measurable DMI values is practically limited by the frequency resolution of tandem Fabry-Perot interferometer, which is an essential part of the BLS technique. The red horizontal dashed line indicates the lower limit of the BLS-measurement scheme.

Therefore, the former two schemes have better sensitivity with weak DMIs, whereas the BLS schemes is better with strong DMIs. These different ranges of ferromagnetic layer thickness and measurable DMI strengths give these measurement schemes mutually complementary roles, once the compatibility between the measurement results have been verified, as demonstrated here.

Comparing the $v_{\mathrm{DW}}$- and $\varepsilon_{\mathrm{ST}}$-measurement schemes, since the additional DW speed asymmetry (or chiral damping) usually disappears in thinner ferromagnetic layers [22], the $v_{\mathrm{DW}}$-measurement scheme is more applicable to ultra-thin (<0.3~0.4 nm) ferromagnetic layers, where the DW speed variation becomes symmetric with respect to a shifted symmetry axis. In this thickness range, a sizeable STT efficiency appears [30], which makes it difficult to analyze the $\varepsilon_{\mathrm{ST}}$ variation based on the SOT theory. This sizeable STT efficiency disappears as the thickness of the ferromagnetic layer increases [30] and thus, the $\varepsilon_{\mathrm{ST}}$ measurement scheme is more applicable to an intermediate (0.5~1.2 nm) range of ferromagnetic layer thickness. The black perpendicular dashed line shows the schematic boundaries for the appearance/disappearance of the chiral damping and the STT efficiency.

The last point to note is that the $v_{\mathrm{DW}}$- and $\varepsilon_{\mathrm{ST}}$-measurement schemes directly quantify the value of $H_{\mathrm{DMI}}$ and thus, it is more useful when the value of $H_{\mathrm{DMI}}$ is required, such as the SOT-induced DW motion, while estimating the DMI requires some additional measurements. On the other hand, the BLS scheme is able to quantify the DMI more rigorously, with only one additional parameter $M_{\mathrm{S}}$ which is easily accessible.

In summary, we investigated the compatibility of different DMI measurement schemes based on two distinct magnetic dynamics, magnetic DW and SW dynamics. After careful measurements and analyses with 6 different magnetic thin films, we found that the measurement results showed fairly good coincidence to each other. This observation removes the possibility that the DMI plays intrinsically different roles in the magnetic DW and SW dynamics, and confirms the reliability of the recently-developed diverse measurement schemes, thus providing a way towards establishing measurement standards.

**ACKNOWLEDGEMENTS / FUNDING SOURCES**

This work was supported by a National Research Foundations of Korea (NRF) grant that was funded by the Ministry of Science, ICT and Future Planning of Korea (MSIP) (2015R1A2A1A05001698, 2015M3D1A1070465, and 2017R1A2B300262). Y.-K. Park and B.-C. Min were supported by the KIST institutional program and Pioneer Research Center Program of MSIP/NRF (2011-0027905). N.-H. Kim, J. Jung, J.-S. Kim and C.-Y. You were supported by the DGIST R&D Programs of the Ministry of Science and ICT (18-NT-01, 18-BT-02, 18-01-HRMA-04, and 18-01-HRMA-01). J. Cho and D.-H. Kim was supported by Postdoctoral Fellowship for Foreign Researchers program (No. P16362) of Japan Society for the Promotion of Science (JSPS).

**AUTHOR CONTRIBUTIONS**

D.-Y. Kim planned and designed the experiment and S.-B. Choe and C.-Y. You supervised the study. D.-Y. Kim, N.-H. Kim, Y.-S. Nam and J.-S. Kim, J. Jung carried out the measurements. M.-H. Park, Y.-K. Park and B.-C. Min prepared the samples. D.-Y. Kim, N.-H. Kim, J. Cho, D.-H. Kim, J.-S. Kim, S.-B. Choe and C.-Y. You performed the analysis. D.-Y. Kim, S.-B. Choe and C.-Y. You wrote the manuscript. All authors discussed the results and commented on the manuscript.

**CORESPONDING AUTHORS**

Prof. Sug-Bong Choe (sugbong@snu.ac.kr) and Prof. Chun-Yeol You (cyyou@dgist.ac.kr)

**Table 1.** $H_\text{K}$, $H_\text{DMI}$, $D_\text{DW}$, $\Delta f$, and $D_\text{SW}$ of the current samples.

| X | $H_\text{K}$ [T] | $H_\text{DMI}$ [mT] | $D_\text{DW}$ [mJ/m$^2$] | $\Delta f$ [GHz] | $D_\text{SW}$ [mJ/m$^2$] |
|---|---|---|---|---|---|
| Ti | 1.13±0.013 | -197±25 | 1.45±0.18 | 2.20±0.26 | 1.52±0.18 |
| Cu | 0.90±0.021 | -190±25 | 1.58±0.19 | 1.57±0.54 | 1.09±0.45 |
| W  | 0.95±0.012 | -183±5  | 1.48±0.03 | 2.15±0.21 | 1.49±0.15 |
| Ta | 1.20±0.018 | -160±10 | 1.15±0.06 | 1.95±0.31 | 1.35±0.21 |
| Al | 0.94±0.049 | -109±5  | 0.88±0.02 | 1.35±0.13 | 0.94±0.09 |
| Pt | 0.79±0.010 | 0±10    | 0.00±0.09 | 0.02±0.13 | 0.02±0.09 |

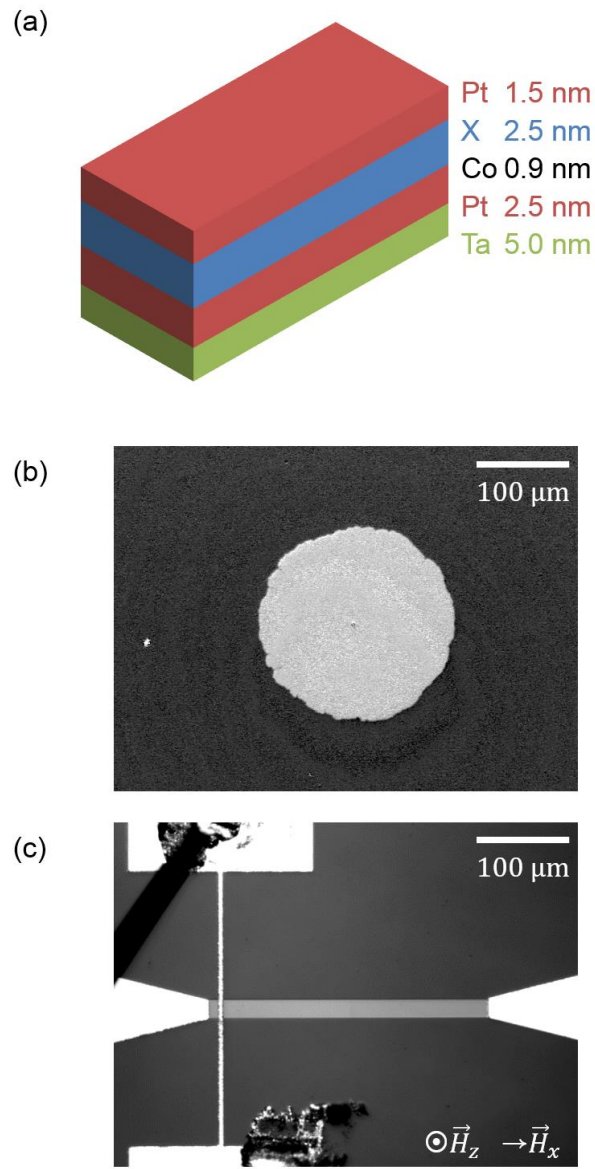

**Figure 1.** (a) Schematic drawing of the multilayer film structure. (b) Magneto-optical Kerr effect (MOKE) image of magnetic domains on continuous film. (c) Optical image of the patterned magnetic micro-wire with electrodes and DW writing line. The light grey horizontal rectangle shows a magnetic wire with a 20 μm width, and the white areas are the 15-nm Ti/100-nm Au electrodes. The white vertical line is the DW writing line.

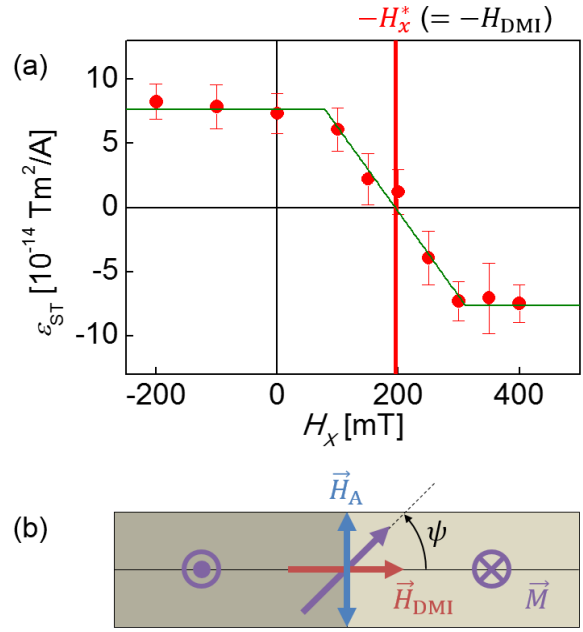

**Figure 2.** (a) Plot of the measured $\varepsilon_{ST}$ as a function of $H_x$. The red vertical line indicates the intercept ($H_x^*$) to the abscissa. The green lines are visual guides to help observe the anti-symmetric nature of the $\varepsilon_{ST}$ variation. (b) Schematic drawing of the DW between two magnetic domains on the micro-wire structure. The purple arrows represent the local magnetizations $\vec{M}$. The red and blue arrows represent the DMI-induced effective field $\vec{H}_{DMI}$ and DW anisotropy field $\vec{H}_A$, respectively. The angle $\psi$ is the angle between $\vec{M}$ and $\vec{H}_{DMI}$.

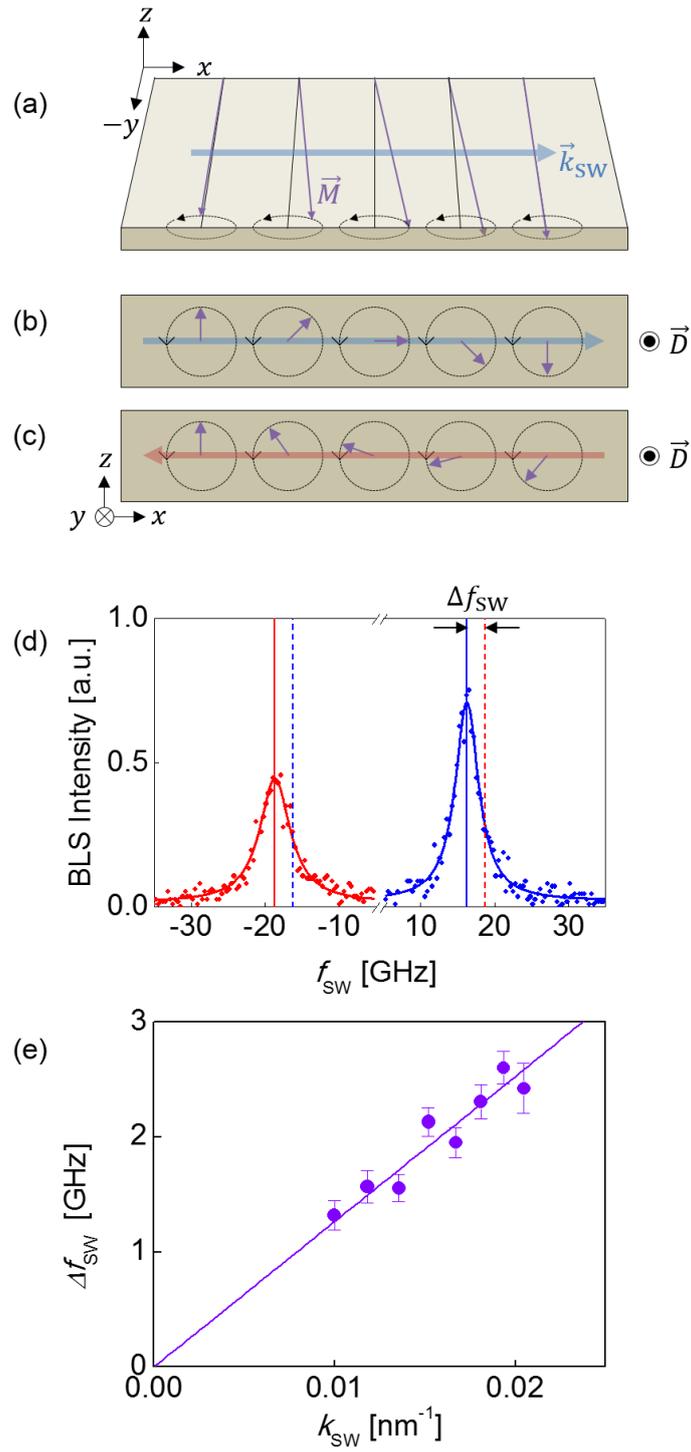

**Figure 3.** (a) Schematic diagram of the DE geometry when a strong $H_y$ is applied along the $-\hat{y}$ direction. The black curved arrows show the direction (counterclockwise) of the procession. The purple arrows show a snapshot of the local magnetization when the SW

propagates along the $+\hat{x}$ direction. The wave vector is shown by the blue arrow. (b), (c) Cross-sectional view of the SWs propagating along the $+\hat{x}$ and $-\hat{x}$ directions, respectively. The purple arrows again show a snapshot of local magnetizations at precession. The wave vectors are shown by the blue and red arrows, respectively. (d) BLS spectrum for the sample with X=Ti. The solid curves show the best Lorentzian fitting. The blue and red solid vertical lines indicate the peaks of the anti-Stokes and Stokes processes, each of which corresponds to the SWs propagating along the $+\hat{x}$ and $-\hat{x}$ directions, respectively. The dashed vertical lines show the values of the two peaks inverted with respect to $f_{SW}=0$, as visual guides for the peak shifts. (e) Plot of the measured $\Delta f_{SW}$ as a function of $k_{SW}$. The solid line is the best linear fitting of the proportionality.

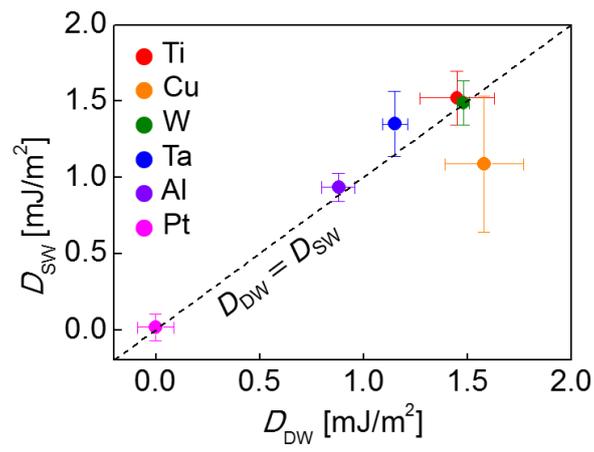

**Figure 4.** Plot of $D_{SW}$ with respect to $D_{DW}$. The dashed line is a visual guide for $D_{DW} = D_{SW}$.

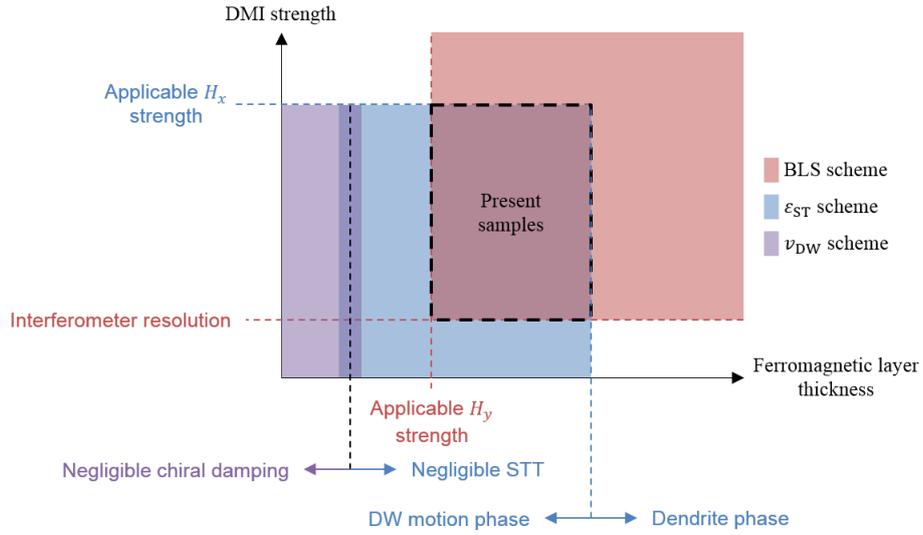

**Figure 5.** Schematic illustration of the recommended measurement schemes with respect to the ferromagnetic layer thickness and DMI strength. The red, blue, and purple shaded boxes show the recommended ranges for the BLS-, $\varepsilon_{ST}$-, and $v_{DW}$-measurement schemes. The applicable range of the BLS-measurement scheme is limited by the interferometer resolution (red horizontal dashed line) and the practically applicable $H_y$ strength (red vertical dashed line). The applicable range of the DW-based schemes (i.e. $\varepsilon_{ST}$- and $v_{DW}$-measurement schemes) is limited by the practically available $H_x$ strength (blue horizontal dashed line) and the phase boundary (blue vertical dashed line) between the DW motion and dendrite phase. The black vertical dashed line shows the applicable boundary between the DW-based schemes due to the appearance/disappearance of the chiral damping and the STT efficiency. The present samples belong to the dashed box area, where both the BLS-measurement scheme and $\varepsilon_{ST}$-measurement scheme are applicable.